 \documentclass[final,5p,times,twocolumn,authoryear]{elsarticle}
\usepackage{amssymb}
\usepackage{amsmath}
\usepackage{graphicx}% Include figure files
\usepackage{dcolumn}% Align table columns on decimal point
\usepackage{bm}% bold math
\usepackage{lipsum}
\usepackage{subfigure}

\usepackage{changes} %使用changes宏包
\usepackage{lineno}

\definechangesauthor[name={Author1}, color=red]{A1} %修订作者1
\definechangesauthor[name={Author2}, color=green]{A2} %修订作者2

\setcitestyle{square,numbers,sort&compress}

\journal{Physics Letters B}

\begin{document}
%\linenumbers



\title{A Toy Model to Explain the Missing Bounce Windows in the Kink-Antikink Collisions}


\author[first]{Lingxiao Long}
\affiliation[first]{organization={School of Space Science and Physics, Shandong University at Weihai},%Department and Organization
            addressline={}, 
            city={Weihai},
            postcode={264209}, 
            state={Shandong},
            country={China}}
\author[first,second]{Xiang Li}

\affiliation[second]{organization={Department of Physics and Astronomy Stony Brook University Stony Brook   },%Department and Organization
            addressline={}, 
            city={NY},
            postcode={11794}, 
            state={},
            country={USA}}
\author[first,third]{Yunguo Jiang\corref{cor1}}
\cortext[cor1]{Corresponding author}
\ead{jiangyg@sdu.edu.cn}
\affiliation[third]{organization={Shandong Provincial Key Laboratory of Optical Astronomy and Solar-Terrestrial Environment Institute of Space Sciences Shandong University at Weihai},%Department and Organization
            addressline={}, 
            city={Weihai},
            postcode={264209}, 
            state={Shandong},
            country={China}}

\begin{abstract}
We propose a toy model to reproduce the fractal structure of kink and antikink collisions on one topological sector of  $\phi^6$ theory. Using the toy model, we investigate the missing bounce windows observed in the fractal structure, and show that the coupling between two oscillation modes modulate the emergence of the missing bounce windows. By numerical calculation, we present that the two bounce resonance corresponds to the lower eigenfrequency of the toy model. 
\end{abstract}
%,Kink, Fractal structure, $\phi^6$ model, Bounce windows
%\keywords{}%Use showkeys class option if keyword
\maketitle
%\tableofcontents

\section{\label{sec:level1}Introduction}
In $\phi^6$ model,the Lagrangian density can be defined as \cite{Gani}
\begin{equation}\label{eqn1} 
\mathcal L=\frac{1}{2}\partial_\mu\phi\partial^\mu\phi-\frac{1}{2}\phi^2(\phi^2-1)^2. 
\end{equation}
Correspondingly, the Euler-Lagrange equation on 1+1 dimension is given as
\begin{equation}\label{eqn1.1} 
\frac{\partial^2 \phi}{\partial t^2}-\frac{\partial^2 \phi}{\partial x^2}+\phi(\phi^2-1)^2+2\phi^3(\phi^2-1)=0.
\end{equation}
The kink soliton\added{s} are non-trivial solutions of Equation (\ref{eqn1.1}), interpolating between two adjacent vacua. We can obtain six soliton configurations by using the BPS bound \cite{bps1,bps2}. Among them, one pair of the kink and antikink (KAK) with configurations $\phi_{(0,1)}=\sqrt{(1+\tanh x)/2}$ and $\phi_{(1,0)}=\sqrt{(1-\tanh x)/2}$ can be obtained. Using the perturbation method, we set $\phi=\phi_s+\eta(x)e^{i\omega t}$, where $\phi_s$ is the static kink configuration.  By linearizing the field equation,  Dorey et al.  presented that \cite{dorey2011kink} 
\begin{equation}\label{eqn2} 
\eta_{xx}=[U(x)-\omega^2] \eta,
\end{equation}
where $U(x)=15\phi_s^4-12\phi_s^2+1$ is the potential. This Schr\"{o}dinger like equation has merely continuous spectrum, which means that any localized vibration mode cannot be found as in the $\phi^4$ model \cite{selfex}.

If we set the kink and antikink with a finite separation and opposite velocities, they would collide and interact with each other, leading to complex phenomena including the resonance and radiation. Numerical simulations demonstrated how the initial velocity and topological sector affect the phenomena of collisions\cite{dorey2011kink}. In the (0,1)+(1,0) sector ($\phi$=$\phi_{(0,1)}(x+a)+\phi_{(1,0)}(x-a)$, \emph{a} represents the half separation between the kink and antikink), the kink and antikink have no resonance. But in the (1,0)+(0,1) sector, one observes the fractal structure \cite{dorey2011kink}. 
%\added{We should note that the mechanism of the formation of the resonance fractal structure between $\phi^4$ and $\phi^6$ theories are subtly different. For $\phi^4$ theory bounce windows are caused by the energy transfer between the translation mode and the localized mode possessed on the single kink, but in $\phi^6$ theory the resonance is triggered by the delocalized modes in the spectrum of small oscillations about a combined kink-antikink configuration.} \cite{dorey2011kink}\added{\cite{Adamrev6,AdamSpe6}}

\deleted{Different with the fractal structure of $\phi^4$ theory, the $\phi^6$ theory shows the "missing" bounce windows (MBW\added{, also called false windows, are the $\nu_{in}$ regimes when they contains the incident velocity, the kink and antikink could separate at a relative large distance after bounces, but for some reason they finally turn back into the bion state}) when $v_{in}$ is between 0.01 and 0.03.} \added{We recall that, in field theory with resonant scattering, $n$-bounce windows are sets of incident velocities when kink and antikink bounce $n$ times before their separation. 3-bounce windows are nested near a 2-bounce window, and for larger $n$ this pattern repeats to compose the fractal structure. In $\phi^4$ theory, the first 2-bounce window occurs with the number of oscillations $n=4$. Then all other 2-bounce windows with higher value $n$ are found at least up to the critical velocity \cite{Manton1}. However, in $\phi^6$ theory, Dorey et al. presented there is one "missing bounce windows" (MBWs), when $n=12$ and $14$ correspond to two 2-bounce windows, but when $n=13$ kink and antikink cannot separate and finally turn into the bion state \cite{dorey2011kink}} \deleted{Such phenomenon is interesting but has no physical explanation. }
To better understand the fractal structure and the MBW phenomena in $\phi^6$ theory, we propose a toy model including two shape modes and one translation mode to reproduce these features and explain the physical mechanism for the emergence of these phenomena. 

\section{The toy model}
%%\label{}
%Owing to the unknown shape mode in $\phi^6$ theory, we invoke the $\phi^4$ theory to present the line of though for the construction of the toy model.  
%In the Collective Coordinate Method (CCM) of the  $\phi^4$ model, the configuration of KAK is a function given as \cite{Manton1}
%\begin{equation}\label{eqn5} 
%\begin{split}
%%%%\end{equation}
%where the parameter \emph{a} is the translation mode, and \emph{b} indicates the shape mode. It is one specified form of KAK configuration for numerical calculation, and it doesn't matter if we replace them with another configuration. 
The essence of CCM is to use dynamics of the translation and shape modes to approximate the partial differential equations (PDEs) in Equation (\ref{eqn1.1})\cite{Manton1}. The spirit of our toy model is to mimic the dynamic of the translation mode by a "particle" passing through a potential well with exponential pattern, and mimic the dynamic of shape mode by an harmonic oscillator. This toy model can successfully reproduce the fractal structure in $\phi^4$ KAK collision, and explain the constraint of the critical velocity  \cite{LIXIANG}. In the numerical results of $\phi^6$ PDE, it is mentioned above that there \deleted{is}\added{are}  missing windows. With more details, the third two-bounce window\added{($\nu_{in}\approx0.033$)} is subtly larger than the neighboring ones, which is absent in the $\phi^4$ model because the width of 2-bounce windows decreases monotonically as \emph{$v_{in}$} \cite{campbell}. 
\begin{figure}[htp]
    \centering
    \includegraphics[width=7cm]{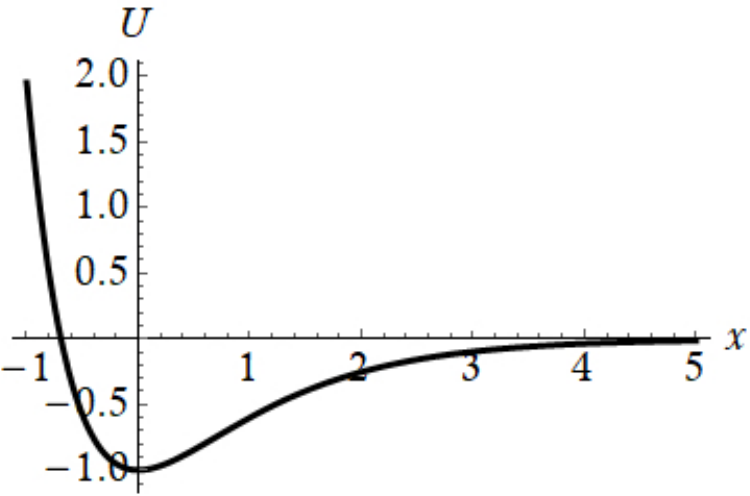}
    \caption{The plot of Morse potential.}
    \label{fig:1}
\end{figure}
\added{In the origin $\phi^6$ model, shape modes exist as the form of bulk modes. From freezing the delocalized modes with the half-separation $a=3$, Adam et al. constructed an effective collective coordinate model (CCM), which reproduced some characteristics in $\phi^6$ theory like critical velocity and 2-bounce windows \cite{Adamrev6}. However, there is no MBW in their model, so they proposed that the missing windows are caused by higher order shape modes or the radiation. That's why we introduced two harmony oscillators to simulate the behaviors of two shape modes. U(a) is chosen to be the Morse potential (See FIG. \ref{fig:1}), which is ideal to describe the interaction between particle-like kink and antikink. When the kink and antikink are far away, they have mutual attraction, and when they pass through each other, they have the strong repulsion.  }\deleted{So, in the toy model of $\phi^6$ theory, we add another harmonic oscillator aiming to reproduce the fractal structure and MBW phenomena.} The Lagrangian of the toy model \added{is consisted of the parts of kinetic energy and the potential, which} can be expressed as
\begin{equation}\label{eqn9}
\begin{split}
{\mathcal L}=&\frac{1}{2}m_1\dot{a}^2+\frac{1}{2}m_2\dot{b}^2+\frac{1}{2}m_3\dot{c}^2+m_{12}\dot{a}\dot{b}+m_{13}\dot{a}\dot{c}\\
&+m_{23}\dot{b}\dot{c}-U(a(t))-U(b(t))-U(c(t)).
\end{split}
\end{equation}
Here $a(t)$ denotes the coordinate of the particle with \added{the Morse} potential $U(a)=e^{-2a(t)}-2e^{-a(t)}$, $b(t)$ and $c(t)$ denote two harmomic oscillator\added{s} with potentials as 
\begin{equation}\label{eqn10}
\begin{split}
&U(b(t))=\frac{1}{2}\omega_1^2 b^2(t), \,\,\,\, U(c(t))=\frac{1}{2}\omega_2^2 c^2(t).
\end{split}
\end{equation}
where $\omega_{1}$ and \replaced{$\omega_{2}$}{$\omega_{1}$}  denote the oscillation frequencies. \added{The coupled terms in the Lagrangian (\ref{eqn9}) infer that energy could transfer between shape modes.}

\added{}

The 
Euler-Lagrangian equations for the three modes are derived as
\begin{equation}\label{eqn13}
\begin{split}
&\Ddot{a}(t)=k_{11}(e^{-2a(t)}-e^{-a(t)})+k_{12}b(t)+k_{13}c(t),\\
&\Ddot{b}(t)=k_{21}(e^{-2a(t)}-e^{-a(t)})+k_{22}b(t)+k_{23}c(t),\\
&\Ddot{c}(t)=k_{31}(e^{-2a(t)}-e^{-a(t)})+k_{32}b(t)+k_{33}c(t).
\end{split}
\end{equation}
Here $k_{ij}$ are explicit functions of coefficients including $m_1,\, m_{12},\, w_1,$  etc.  

We study the dynamic of the toy model with numerical calculation. To observe the fractal structure, we show the numerical output of the field
\begin{equation}\label{eqn11}
\begin{split}
\phi(x;a(t))=\phi_{(0,1)}(x-a(t))+\phi_{(1,0)}(x+a(t)).
\end{split}
\end{equation}
We don't consider shape modes in Equation (\ref{eqn11}) because parameters \emph{b(t)} and \emph{c(t)} have ignorable contribution to the fractal structure. We observe that when the kink and antikink are separated far enough, the evolution of \emph{b(t)} and \emph{c(t)} is independent of \emph{a(t)}. It is obvious that the second and third equations in (\ref{eqn13}) constitute a set of coupled oscillators equations. It means that the system have two  independent eigenfrequencies \cite{LANDAU1}, which can be derived by the decoupling technique as
\begin{subequations}
\label{eq:whole}
\begin{equation}\label{subeq:2}
\begin{split}
\tilde\omega_1=&(-m_{13}^2 \omega _1^2+m_1 m_3 \omega _1^2-m_{12}^2 \omega _2^2+m_1 m_2 \omega _2^2\\&+(m_{13}^4 \omega _1^4+m_1^2 m_3^2 \omega _1^4-2 m_1
   m_3 m_{13}^2 \omega _1^4\\&+2 m_1 m_3 m_{12}^2 \omega _2^2 \omega _1^2+2 m_{12}^2 m_{13}^2 \omega _2^2 \omega _1^2\\&+2 m_1 m_2 m_{13}^2 \omega _2^2
   \omega _1^2+4 m_1^2 m_{23}^2 \omega _2^2 \omega _1^2\\&-2 m_1^2 m_2 m_3 \omega _2^2 \omega _1^2-8 m_1 m_{12} m_{13} m_{23} \omega _2^2 \omega
   _1^2+m_{12}^4 \omega _2^4\\&+m_1^2 m_2^2 \omega _2^4-2 m_1 m_2 m_{12}^2 \omega _2^4)^\frac{1}{2})/(-2 m_3 m_{12}^2\\&+4 m_{13} m_{23} m_{12}-2 m_2 m_{13}^2-2 m_1
   m_{23}^2+2 m_1 m_2 m_3),
\end{split}
\end{equation}
\begin{equation}\label{subeq:1}
\begin{split}
\tilde\omega_2=&(-m_{13}^2 \omega _1^2+m_1 m_3 \omega _1^2-m_{12}^2 \omega _2^2+m_1 m_2 \omega _2^2\\&-(m_{13}^4 \omega _1^4+m_1^2 m_3^2 \omega _1^4-2 m_1
   m_3 m_{13}^2 \omega _1^4\\&+2 m_1 m_3 m_{12}^2 \omega _2^2 \omega _1^2+2 m_{12}^2 m_{13}^2 \omega _2^2 \omega _1^2\\&+2 m_1 m_2 m_{13}^2 \omega _2^2
   \omega _1^2+4 m_1^2 m_{23}^2 \omega _2^2 \omega _1^2\\&-2 m_1^2 m_2 m_3 \omega _2^2 \omega _1^2-8 m_1 m_{12} m_{13} m_{23} \omega _2^2 \omega
   _1^2+m_{12}^4 \omega _2^4\\&+m_1^2 m_2^2 \omega _2^4-2 m_1 m_2 m_{12}^2 \omega _2^4)^\frac{1}{2})/(-2 m_3 m_{12}^2\\&+4 m_{13} m_{23} m_{12}-2 m_2 m_{13}^2-2 m_1
   m_{23}^2+2 m_1 m_2 m_3).
\end{split}
\end{equation}
\end{subequations}
%To distinguish these different types of frequencies, we denote  eigenfrequency as "\emph{$\tilde\omega$}" and frequency in numerical results as "\emph{$\overline\omega$}". We see that $\tilde{\omega}$ has little difference to corresponding $\omega$, because all coupled coefficients($m_{12},\,m_{13},\,m_{23}$) are in the perturbative interval.
\section{Numerical result}
%%\label{}
%first. 
We use the "Ndsolve" in {\it Mathematica} to perform the numerical calculation.
From the Lagrangian in Equation (\ref{eqn9}), the parameter $m_{23}$ can effect the intensity of energy transfer between two oscillation modes. 
Initially, We set $m_1=m_2=m_3=1$ and $m_{12}=m_{13}=0.1$. This ensures that the two oscillators are equivalent when we ignore the frequency. The initial half separation is set as $a(0)=5$ across this work. 

\begin{figure}[htp]
    \centering
    \includegraphics[width=9cm]{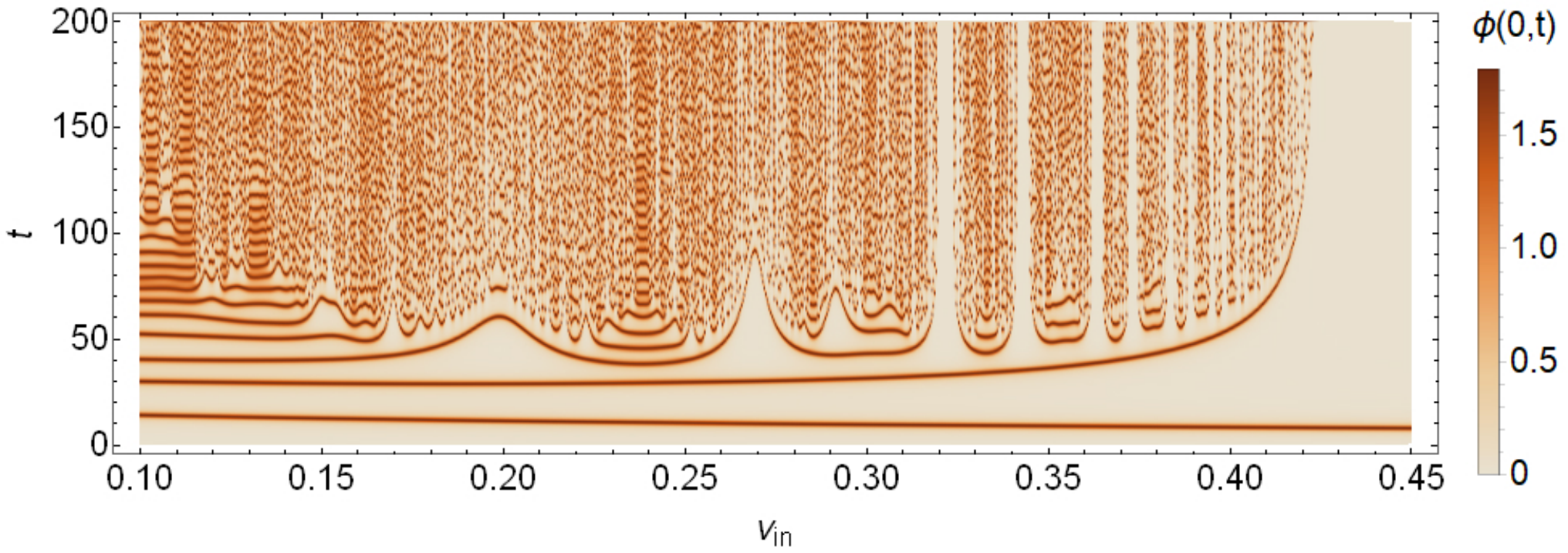}
    \caption{The plot of $\phi(0,t)$ for the toy model. $\nu_{in}$ denotes the incident velocity. The initial parameters are set as $\omega_1=2,\omega_2=3,m_{23}=0$.}
    \label{fig:2}
\end{figure}
 First, we fix $\omega_1=2$, $\omega_2=3$, and try different settings of  $m_{23}$ to show its role in modulating the MBW. 
In FIG. \ref{fig:2}, we first plot the  $m_{23}=0$ case. For \replaced{$0.32< \nu_{in}<0.34$}{$0.32< V_{in}<0.4$}, the regime between two adjacent two bounce windows, namely the 'three bounce regime', show two different patterns appearing alternatively.  The non-monotonicity of the width\added{s} of two-bounce windows may be correlated with the dual patterns.  

\begin{figure}[htbp]
    \centering
    \subfigure[$m_{23}=0.05$]{
        \includegraphics[width=9cm]{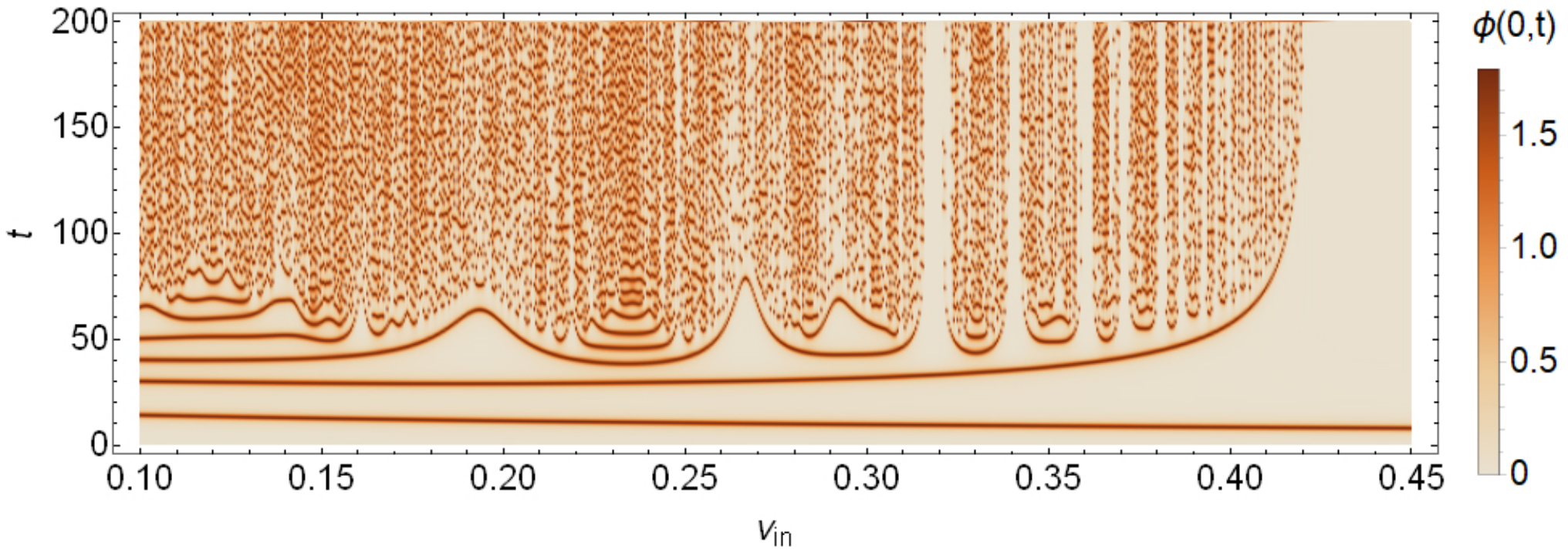}
        \label{label_for_cross_ref_1}
    }
      %用 \quad 来换行
    \subfigure[$m_{23}=0.10$]{
    	\includegraphics[width=9cm]{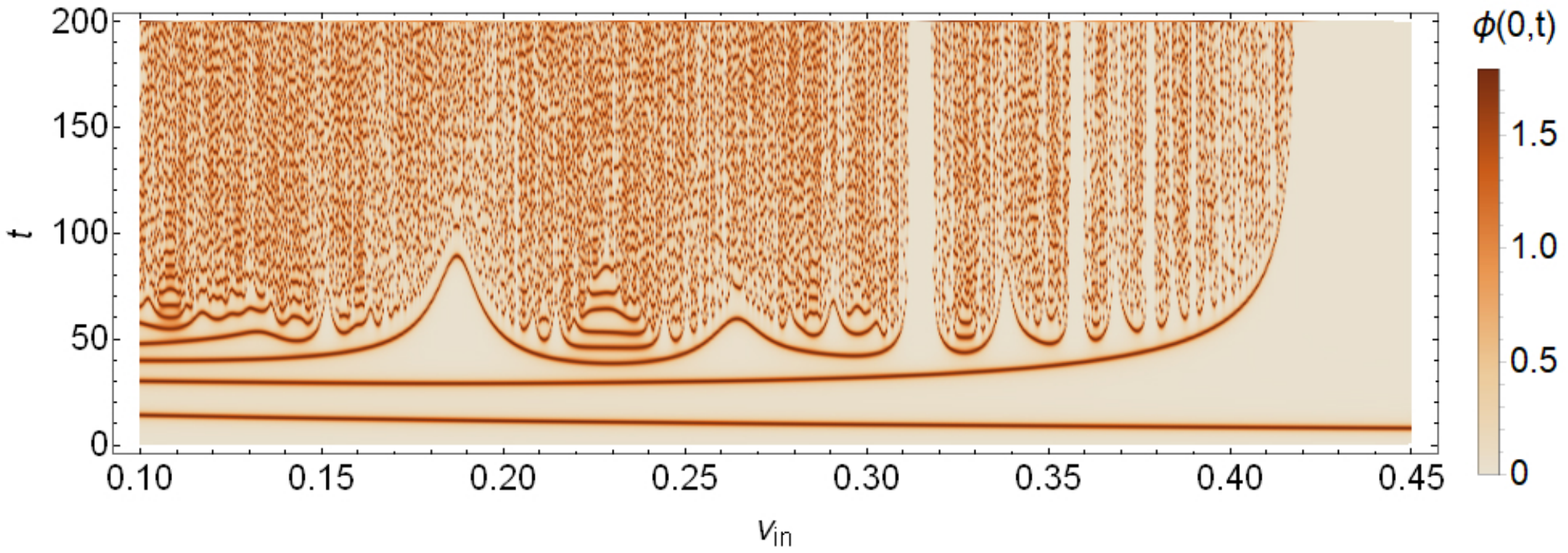}
        \label{label_for_cross_ref_2}
    }
    \quad    %用 \quad 来换行
    \subfigure[$m_{23}=0.15$]{
    	\includegraphics[width=9cm]{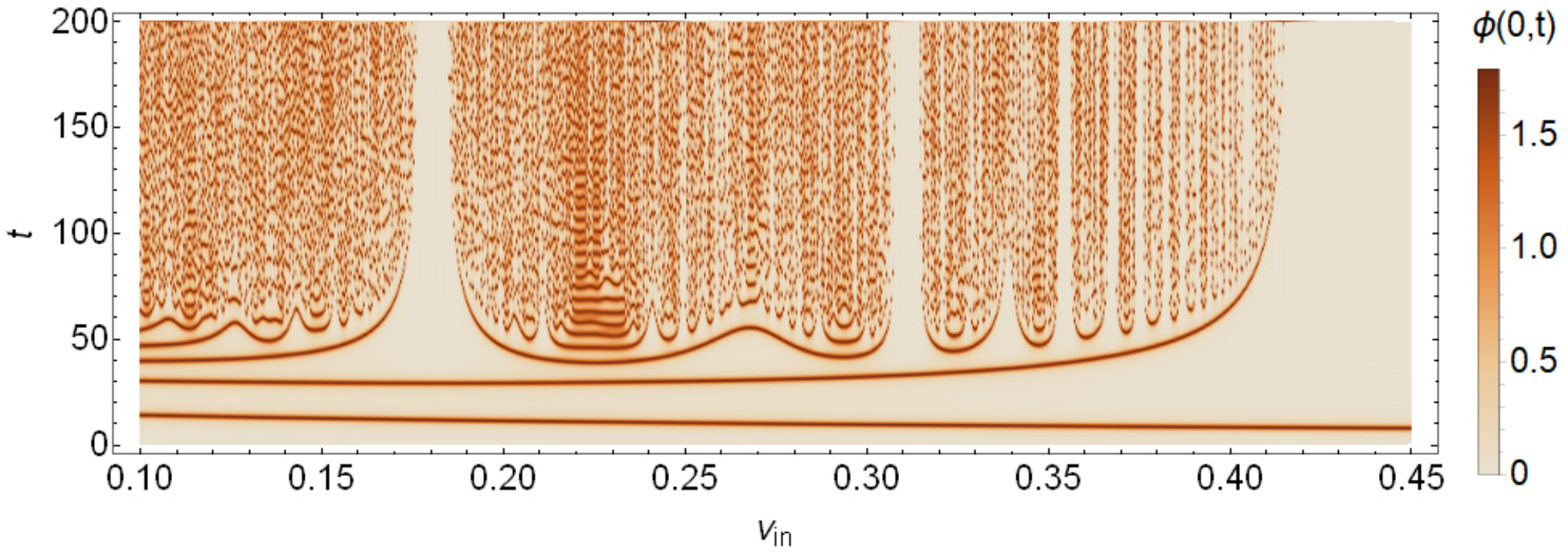}
        \label{label_for_cross_ref_3}
    }
      %用 \quad 来换行
    \subfigure[$m_{23}=0.20$]{
    	\includegraphics[width=9cm]{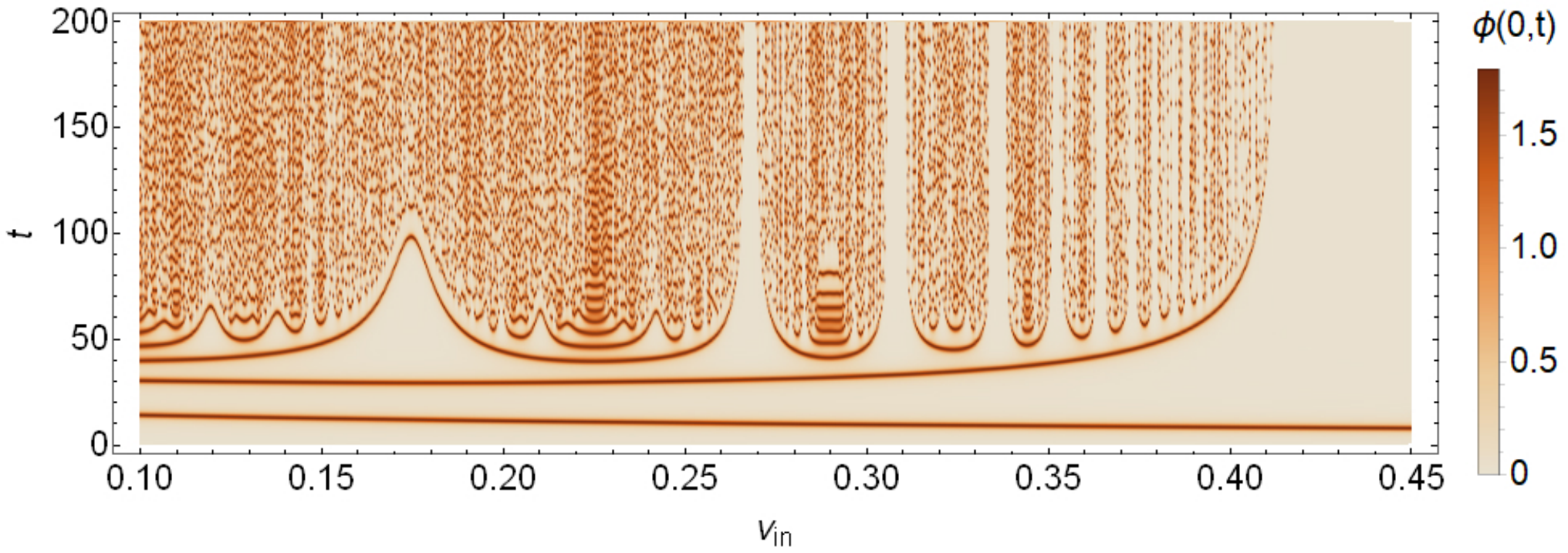}
        \label{label_for_cross_ref_4}
    }
    \quad    %用 \quad 来换行
    \subfigure[$m_{23}=0.25$]{
    	\includegraphics[width=9cm]{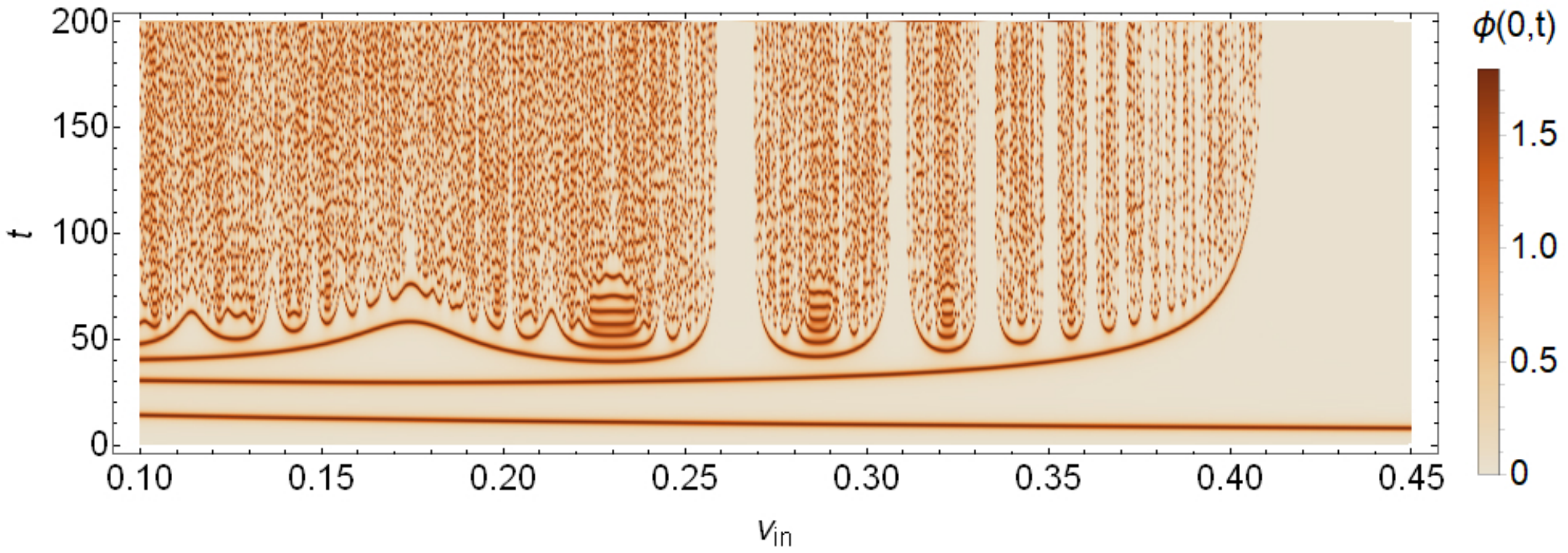}
        \label{label_for_cross_ref_5}
    }
    \caption{Plots of the center field value $\phi(0,t)$ as a function of initial velocity and time in the toy model with different  $m_{23}$ ($\omega_1=2$,\,$\omega_2=3$,\,$a_0=5$).}
    \label{fig.3}
\end{figure}

In FIG. \ref{fig.3}, we plot the the center field value $\phi(0,t)$ \added{as a function of the time and incident velocities} for $m_{23}=0.05, 0.10, 0.15, 0.20, 0.25$ in panels (a), (b), (c), (d), and (e), respectively.
When $m_{23}=0$ (see FIG. \ref{fig:2}), there are two MBWs near $v_{in}=0.20$ and $v_{in}=0.27$. As $m_{23}$ increases to 0.10, the first MBW gets deeper (If the third collision time of a MBW gets longer, we call that the MBW becomes deeper, inversely we call it shallower) while the second one becomes shallower. Meanwhile, as seen in FIG. \ref{label_for_cross_ref_2}, the escaping window at $ V_{in}\sim 0.34$  turns to a MBW. The current first MBW \replaced{has}{have} 'penetrated the ceiling' and \replaced{converts}{converted} into a resonance window when $m_{12}$ equals to 0.15. It can also be observed that some MBWs get deeper and some turn back into resonance windows. When $m_{23}$ reaches 0.20 (see FIG. \ref{label_for_cross_ref_4}), the window near $v_{in}=0.18$ turn\added{s} back into a MBW. In the meantime, these appeared MBWs for $0.32< V_{in}<0.4$ vanish again. \replaced{The fractal}{Fractal} structure tends to be stable as $m_{23}$ gets enough larger but less than 1 (FIG. \ref{label_for_cross_ref_5}). We observed that the coefficient $m_{23}$ has ignorable effect on the critical velocity, but it can effect \deleted{significantly }the fractal structure \added{significantly observed }from FIG. \ref{fig.3}. It also plays an important role in the formation of two-bounce windows. 

Another difference between the $\phi^6$ model and $\phi^4$ model is that more than one oscillation modes are contained in the former theory. Although there \replaced{is}{are} no local self-excitation, recent literature showed that delocalized oscillation modes do exist with different frequencies \cite{Adam}. It is widely known that when resonance scattering exists, \added{there are favorable timings where the shape mode would be excited by the second impact through some characteristic phase angle. }\added{Thus, the resonance condition can be expressed as}\deleted{the resonance frequency could be detected in the resonance condition}\cite{campbell}:
\begin{equation}\label{eqn14}
\omega_sT=2n\pi +\delta.
\end{equation}
Where \emph{T} is the time interval between two collisions, and \emph{n} represents the number of lower frequency oscillation in the period. $\delta$ is the phase position\cite{campbell}. 

We propose that the resonance frequency may \replaced{relate}{related} to the eigenfrequencies. To \replaced{test}{testing} our proposal, we plot the $v_{in}$ of the centres of each 2-bounce windows versus the oscillation \replaced{number of times}{time} $n$ from\added{ the} numerical calculation. We don't count for the start number of bounce in the first window. So, when depicting scattering points, $n$  always starts at 1. 

\begin{figure}[htbp]
    \centering
        \subfigure[$\omega_1=3$,\,$\omega_2=4$,\,$m_{23}=0.2$]{
    	\includegraphics[width=8cm]{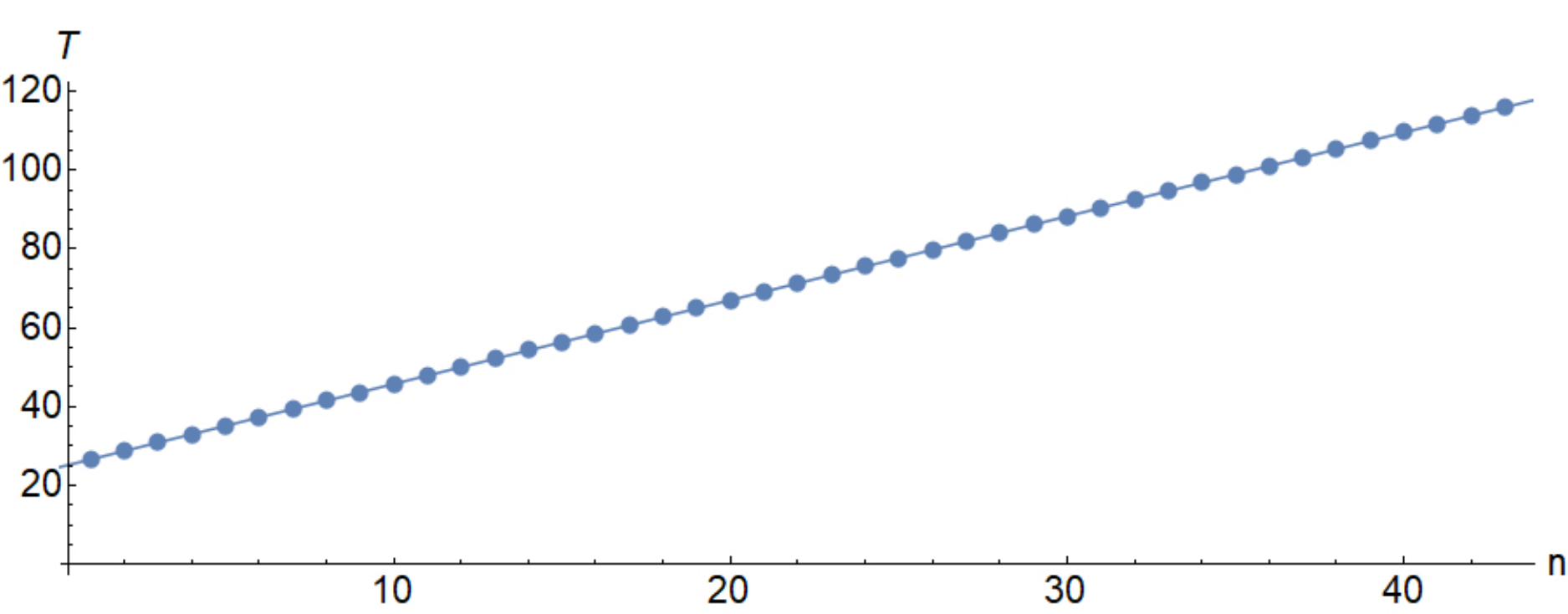}
        \label{51}
    }
    \quad    %用 \quad 来换行
    \subfigure[$\omega_1=5$,\,$\omega_2=2$,\,$m_{23}=0.4$]{
        \includegraphics[width=8cm]{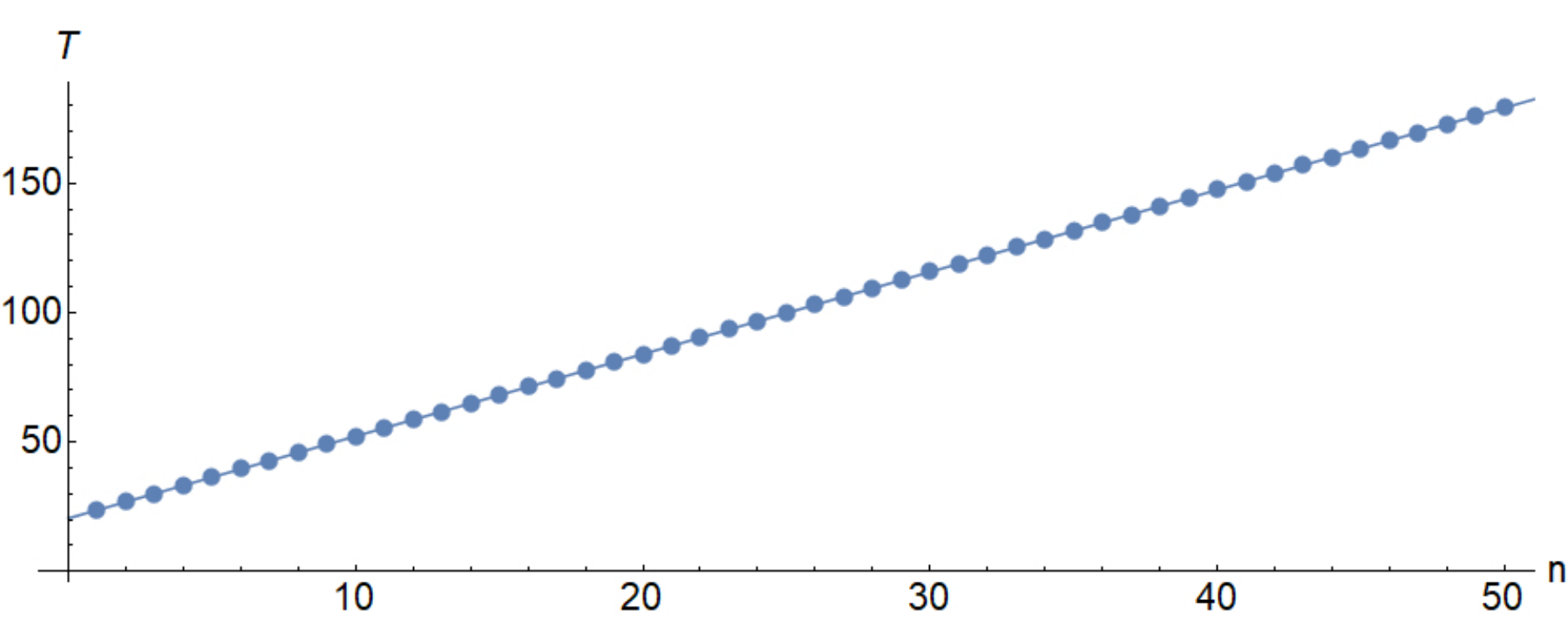}
        \label{52}
    }
    \quad  %用 \quad 来换行
    \caption{\emph{T} versus $n$ for two sets of ($\omega_1,\,\omega_2,\,m_{23}$) are plotted in the top and bottom panels, respectively. The straight lines are obtained by using the least square method.}
    \label{fig.6}
\end{figure}
In the top panel of FIG. \ref{fig.6}, the slope of fitted straight is $2\pi/\,{\overline{\omega}}$, where $\overline{\omega}=2.9522$ agrees exactly with $\tilde\omega$=2.9522 from Equation (\ref{eq:whole}). While in the bottom panel, $\overline{\omega}$=1.9813 is also very close to the $\tilde\omega$=1.9819. In Table \ref{tab1}, we list pairs of $\overline{\omega}$ and  $\tilde\omega$ for different $m_{23}$ to exhibit their correlation. Thus, both two values are close to the minimal frequencies for each case. We conclude that the whole resonance system always responds to the minimum eigenfrequency, since the lower frequency needs lower energy to be excited. \deleted{The higher order mode is important to produce the MBWs in our toy model, which was mentioned as a conjecture in \cite{rev6}.}
\begin{figure}[htbp]
    \centering
        \subfigure[$m_{23}=0.04$]{
    	\includegraphics[width=8cm]{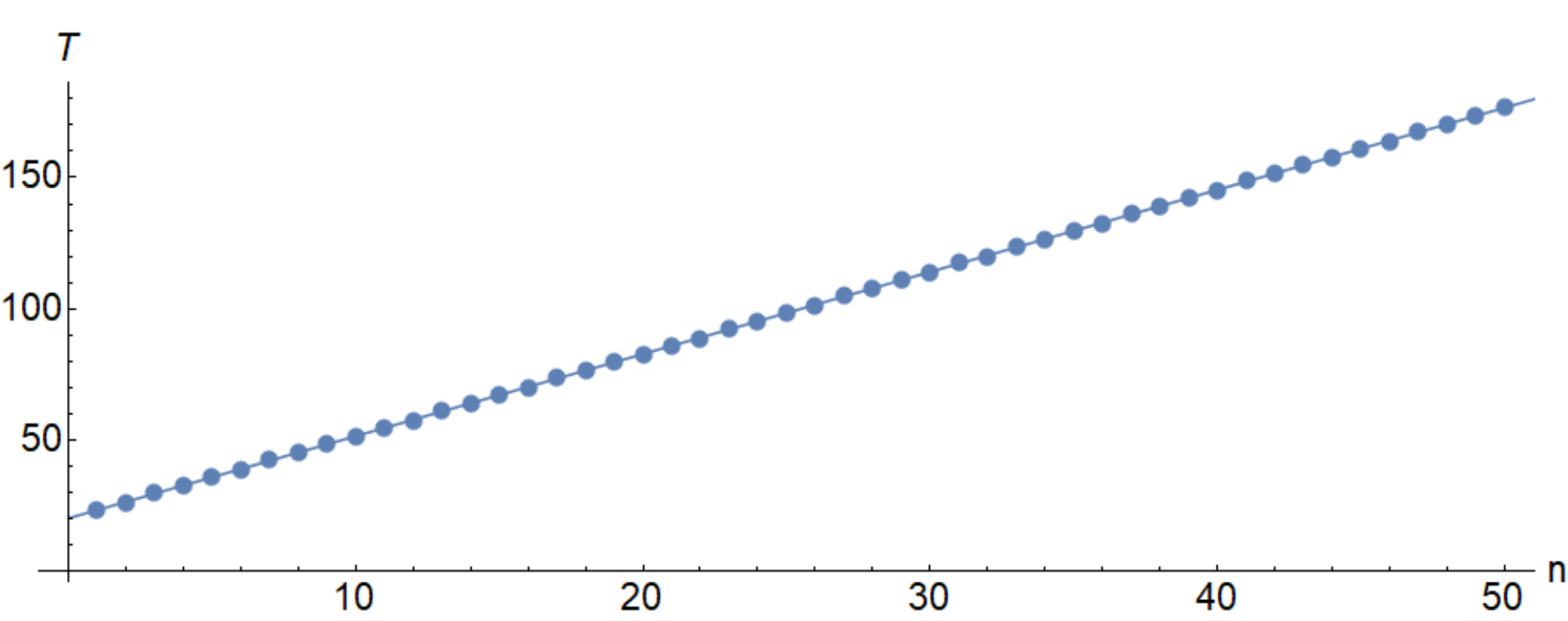}
        \label{71}
    }
        \quad    %用 \quad 来换行
         \subfigure[$m_{23}=0.06$]{
    	\includegraphics[width=8cm]{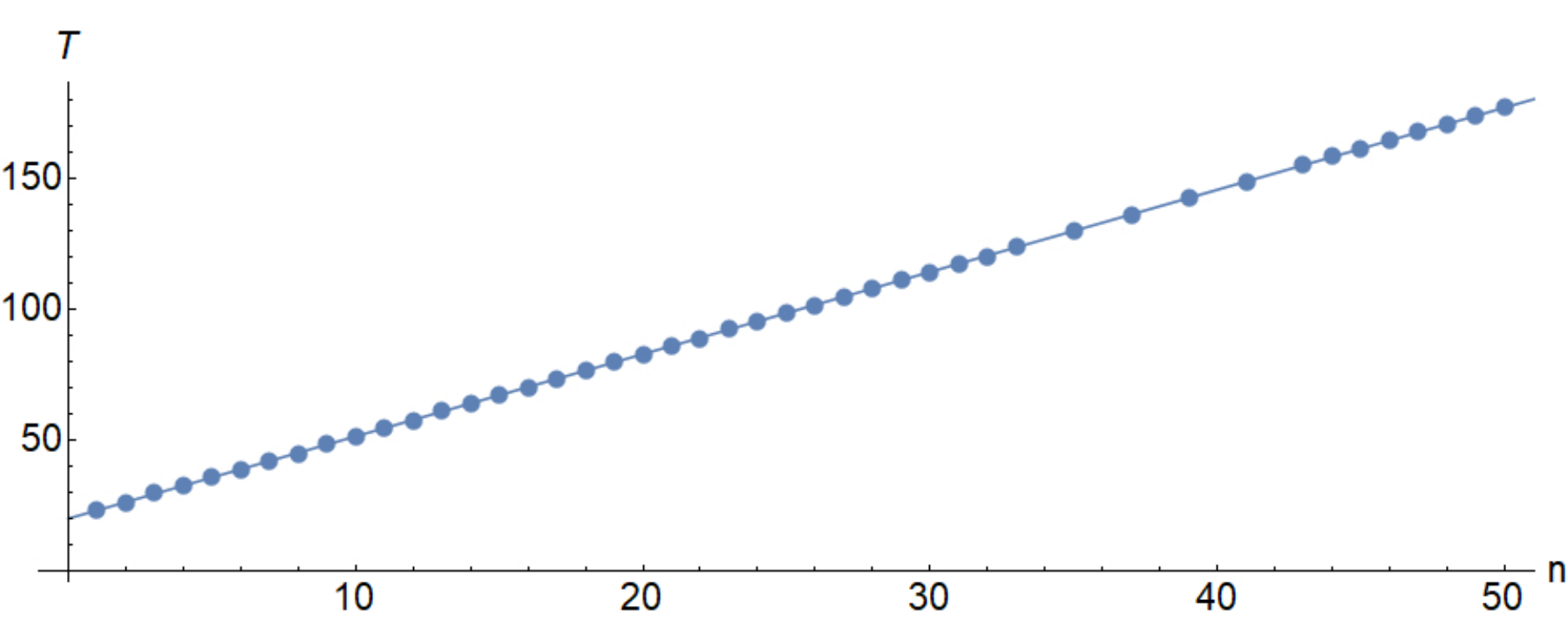}
        \label{72}
    }
        \quad    %用 \quad 来换行
    \subfigure[$m_{23}=0.07$]{
        \includegraphics[width=8cm]{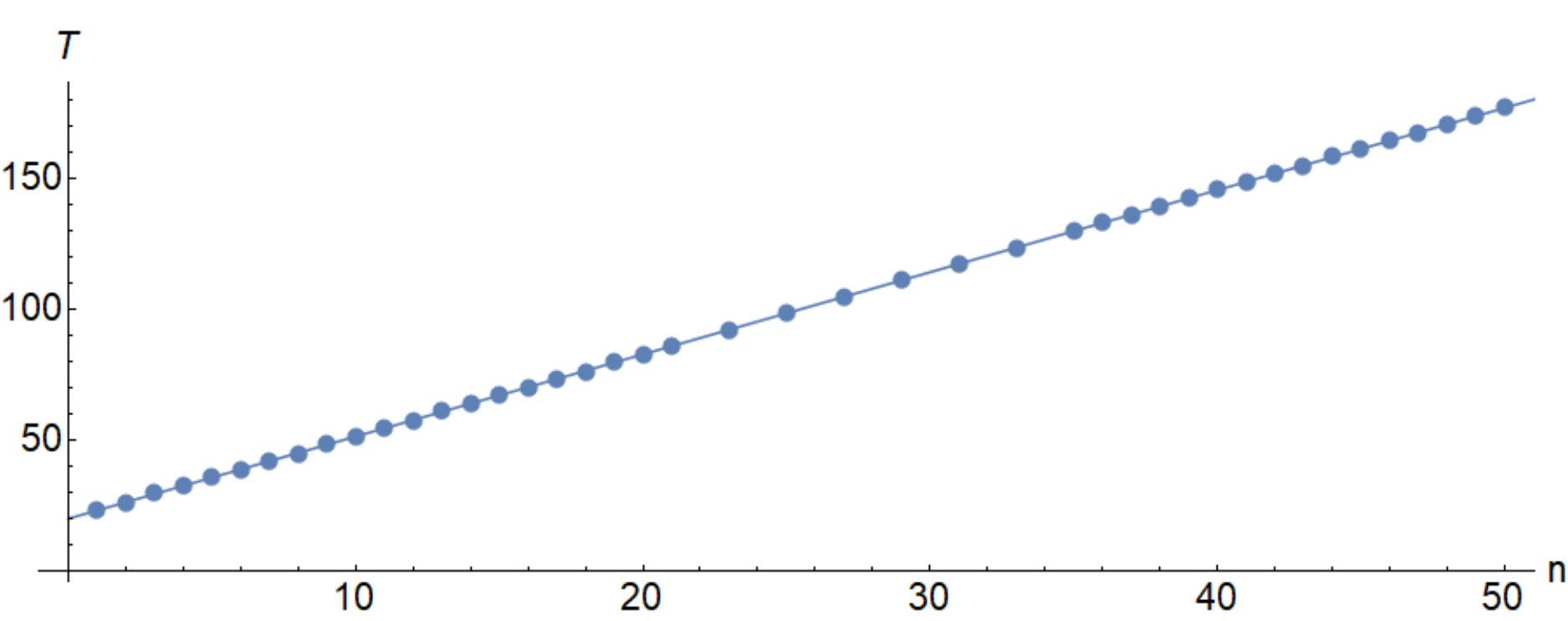}
        \label{73}
    }
    \quad  %用 \quad 来换行
    \subfigure[$m_{23}=0.08$]{
    	\includegraphics[width=8cm]{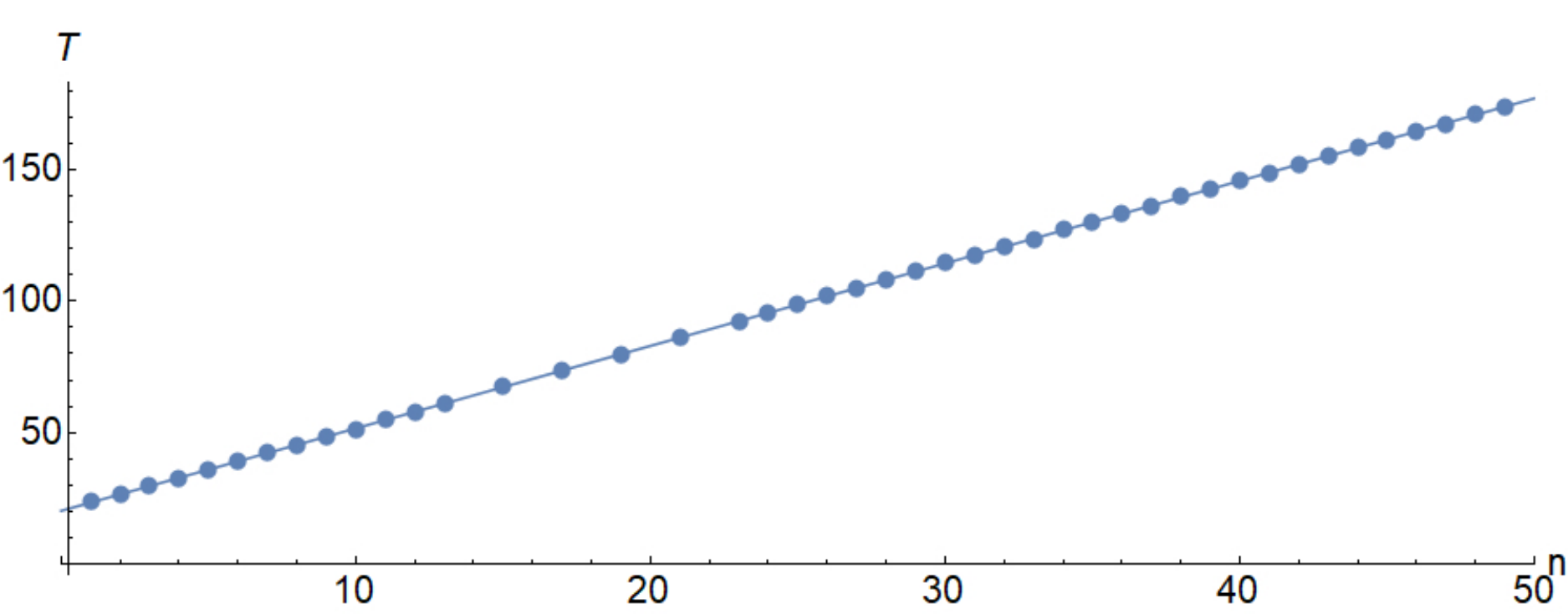}
        \label{74}
    }
        \quad    %用 \quad 来换行
    \subfigure[$m_{23}=0.11$]{
    	\includegraphics[width=8cm]{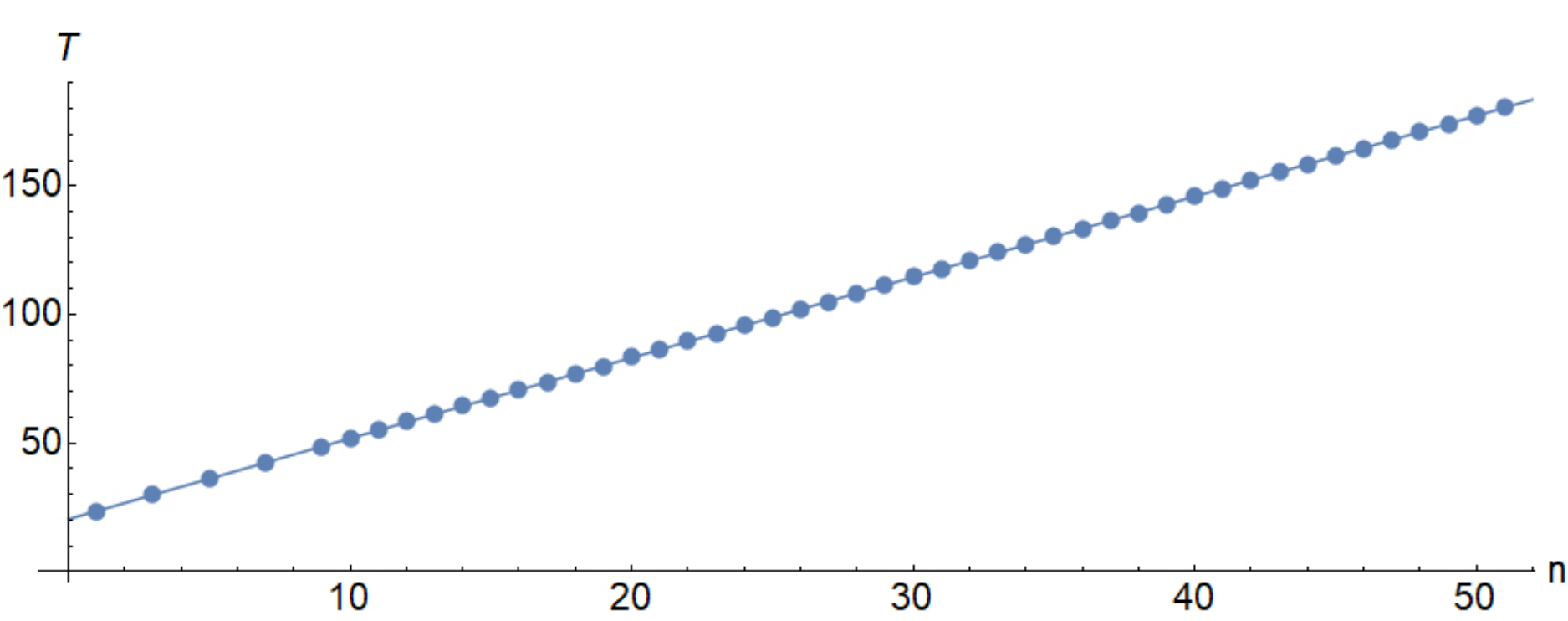}
        \label{75}
    }
    \quad    %用 \quad 来换行
    \subfigure[$m_{23}=0.13$]{
    	\includegraphics[width=8cm]{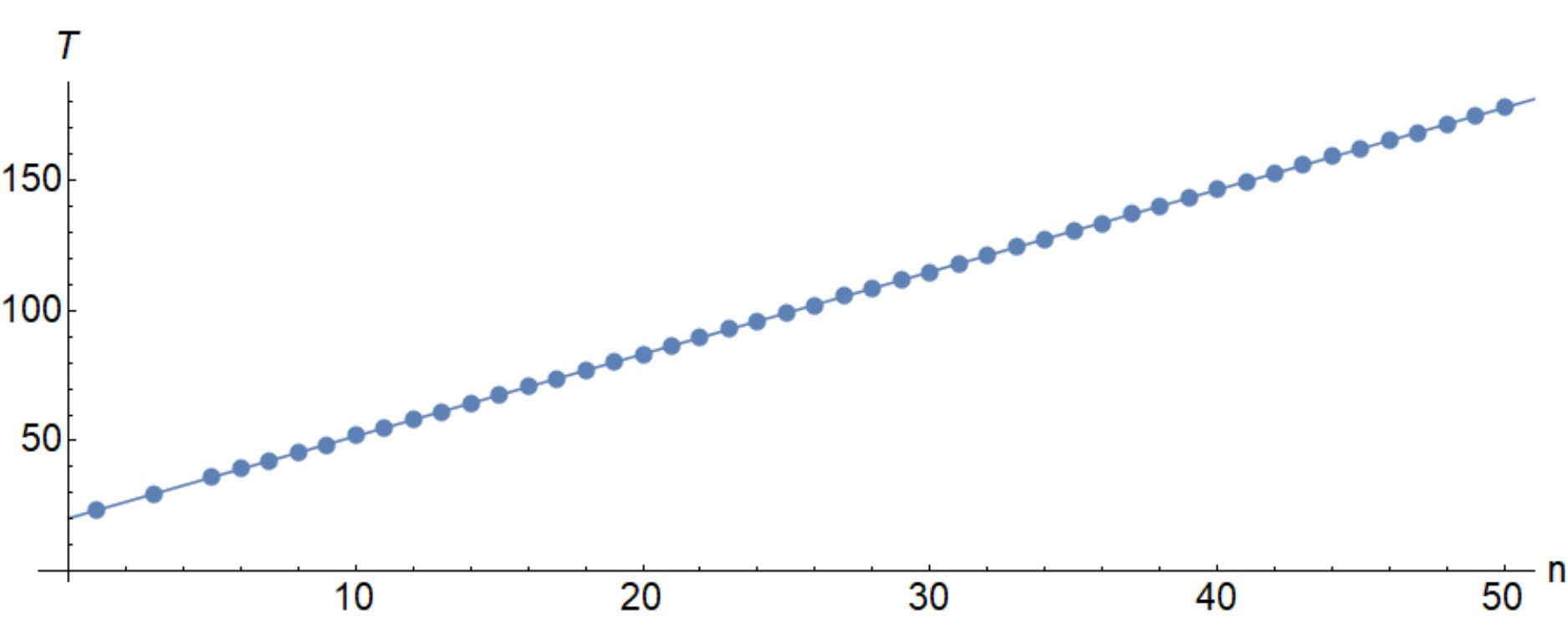}
        \label{76}
    }
        \quad    %用 \quad 来换行
    \caption{\emph{T} as a function of \emph{n} at centres of 2-bounce windows for different $m_{23}$ when $\omega_1=2,\,\omega_2=3$.}
    \label{fig.7}
\end{figure}

\begin{table}
\begin{tabular}{c c c c} 
 \hline
$m_{23}$ & $\overline{\omega}$ & $\tilde{\omega}_{min}$ & $\left|\overline{\omega}-\tilde{\omega}_{min}\right|$ \\
 \hline
0.04 & 2.01007 & 2.00934 & 0.00073\\
0.06 & 2.00652 & 2.00804 & 0.00152\\
0.07 & 2.00436 & 2.00714 & 0.00278\\
0.08 & 2.00426 & 2.00610 & 0.00184\\
0.11 & 2.00223 & 2.00204 & 0.00019\\
0.13 & 1.99746 & 1.99860 & 0.00114\\
 \hline
\end{tabular}
\caption{A tabulation of $\overline{\omega}$ and the minimum of two eigenfrequeies etc. for different $m_{23}$ when $\omega_1=2,\,\omega_2=3$. The fourth column represents the absolute value of the difference between $\overline{\omega}$ and the lower value eigenfrequency.
}
\label{tab1}
\end{table}

Now we study where missing phenomenon could exist. we fix $\omega_1=2$, $\omega_2=3$ again, and plot the time between the two collisions versus \emph{n} by using the centres of 2-bounce windows for $m_{23}$ equals to 0.04, 0.06, 0.07, 0.08, 0.11 and 0.13. In FIG. \ref{fig.7}, we show these plots in panels (a), (b), (c), (d), (e) and (f), respectively. There is no missing phenomenon in the first fifty 2-bounce windows when $m_{23}$=0.04. When $m_{23}$ is 0.06, the 34th, 36th, 38th, 40th and 42th bounce numbers correspond to MBWs. Setting $m_{23}=0.07$, MBWs appear when $n=22,24,26,28,30,32,34$. Furthermore, when $m_{23}=0.08$ and $0.11$, these numbers are $n=15,17,19,21,23$ and $n=2,4,6,8$, respectively. Ultimately, only $n=2$ and $n=4$ correspond to the two MBWs. 
%From \ref{fig.7}, we observed that MBWs are always adjacent to each other whenever missing phenomenon appears. We call the set of responding \emph{n} of MBWs missing zone. The zone moves forward When $m_{23}$ becomes larger. We can see when it is 

\begin{figure}[htbp]
    \centering
        \subfigure[the first 2-bounce window when $v_{in}=0.1805$.]{
    	\includegraphics[width=8cm]{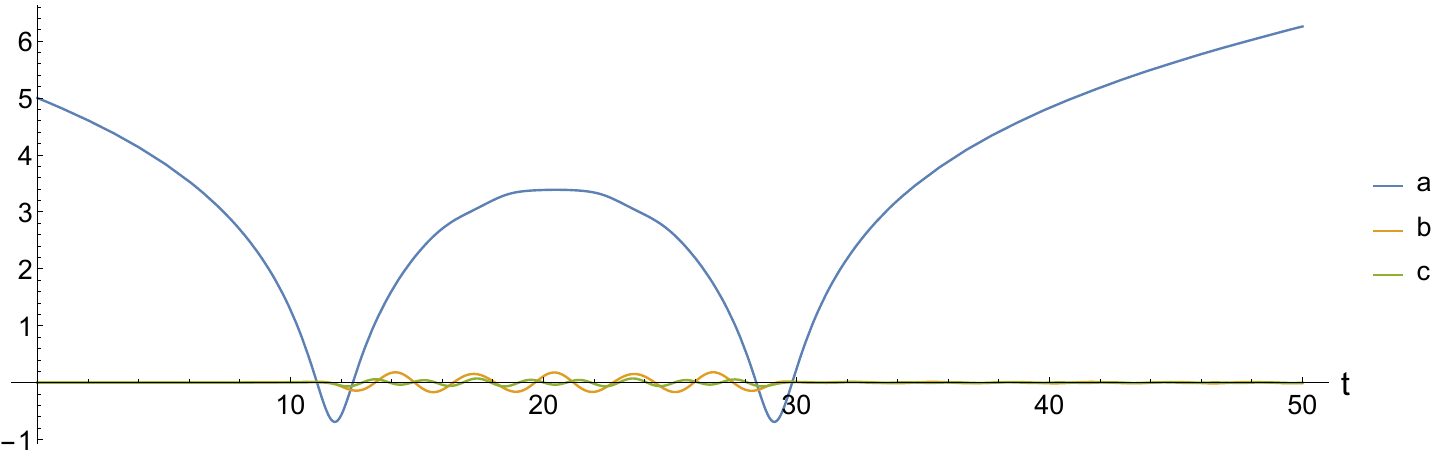}
        \label{81}
    }
    \quad    %用 \quad 来换行
    \subfigure[the first MBW when $v_{in}=0.266$.]{
        \includegraphics[width=8cm]{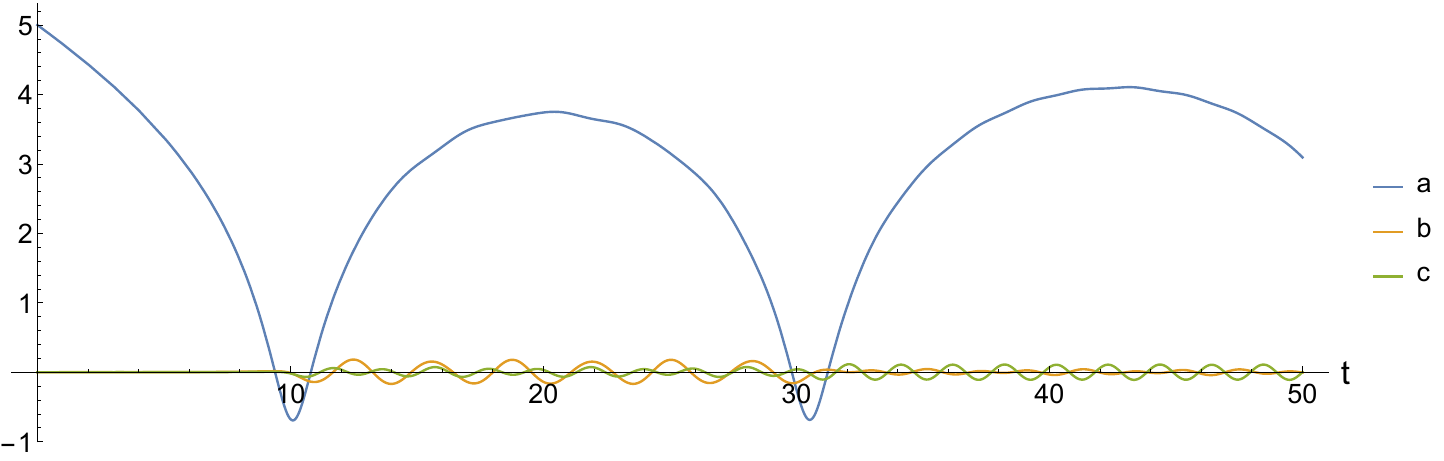}
        \label{82}
    }
    \quad  %用 \quad 来换行
        \subfigure[the second 2-bounce window when $v_{in}=0.311$.]{
        \includegraphics[width=8cm]{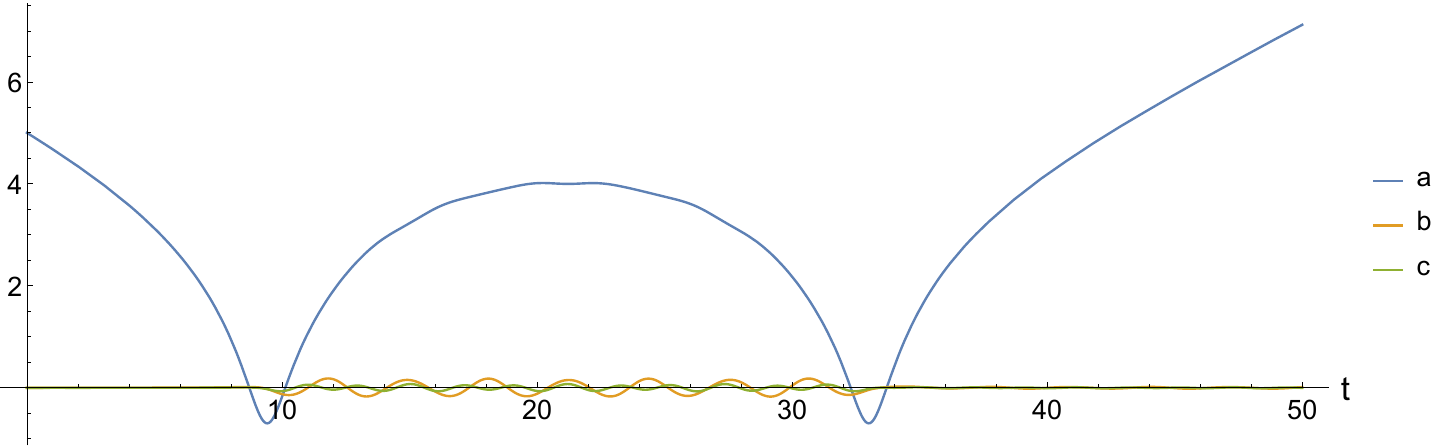}
        \label{83}
    }
    \quad
    \caption{Value of parameters \added{\emph{a},} \emph{b} and \emph{c} as functions of time when $\omega_1=2$,\,$\omega_2=3$,\,$m_{23}=0.15$ (See FIG.\ref{label_for_cross_ref_3}).\deleted{ In panel (a) and (c),  the energy is  almost completely transferred from the oscillation modes to the translation mode. In panel (b), the energy is transferred between two oscillation modes, so that the kink and antikink have not enough kinetic energy to reach the infinity.}}
    \label{fig.8}
\end{figure}

The appearance of missing windows observed in FIG. \ref{fig.7} is most probably due to that there is energy transfer between two oscillation modes. We set $\omega_1=2,\,\omega_2=3,m_{23}=0.07$ (FIG. \ref{73}) for illustration. When \emph{n} is larger than 34, although the energy transfer may still exist somewhere, the total energy is large enough that the third collision wouldn't happen. On the other hand, when \emph{n} is less than 22, there is not enough time to support an effective energy transfer between oscillation modes. That can explain that the missing windows shifts towards  \emph{n}=0 as $m_{23}$ rises. In FIG. \ref{fig.8}, we list the plots of \added{\emph{a(t)}, }\emph{b(t)} and \emph{c(t)} for 2-bounce windows and a MBW to show the dynamics. \added{In FIG. \ref{fig.8}, panel (a) and (c), the energy is almost completely transferred from the oscillation modes to the translation mode. In panel (b), the energy is transferred between two oscillation modes, so that the kink and antikink have not enough kinetic energy to reach the infinity.}%We conjecture that there is only one missing zone and there is just one missing 2-bounce window in $\phi^6$ model.

\added{The equivalence of the frequency (fitted) and the lowest eigenfrequency is not a coincidence. In our toy model, we take $\omega_1=2$,\,$\omega_2=3$,\,$m_{23}=0.15$ for illustration. We show the dynamics for the first 2-bounce window in panel (a) of FIG. \ref{fig.8}. It can be seen that the maximum half-distance between the two collisions is $a_{max}\approx 3.2$ and $n=5$. In panel (b) and (c),  we can see that  $a_{max}$ increases as $n$ been larger.  When $a_{max}$ is larger, the Morse potential disappears, and the eigenfrequency plays a prominent role.  Thus, the eigenfrequency  becomes the fitted frequency for large $n$ limit. }

\added{In $\phi^6$ theory, there are four delocalized modes  at the half separation $a\approx6$ \cite{dorey2011kink}. The fitted frequency $\omega=1.0452$ from the plot of $T$ versus $n$   equals to the frequency of the first delocalized mode. 
These delocalized modes depends on the value of $a$.  In cases of 2-bounce resonance scattering, $a\approx6$ is the maximum half-distance that kink and antikink could reach in the period of the first and the second impact. 
Thus, the "eigenfrequency" in the $\phi^6$ model may be  the lowest one of the four frequencies of delocalized modes at  $a=6$. 
}

\added{In our toy model, the two frequencies of oscillators are different, and the lower one dominates the 2-bounce windows. The higher frequency is evident in the MBW (see panel (b) in Fig R4, where the green line denotes the dynamics of the coordinate ‘c’). Thus, our toy model illustrates that the MBW are the result of multiple eigenfrequencies, and sheds light on the MBW mechanism in $\phi^6$ model.}

\begin{figure}[htp]
    \centering
    \includegraphics[width=8cm]{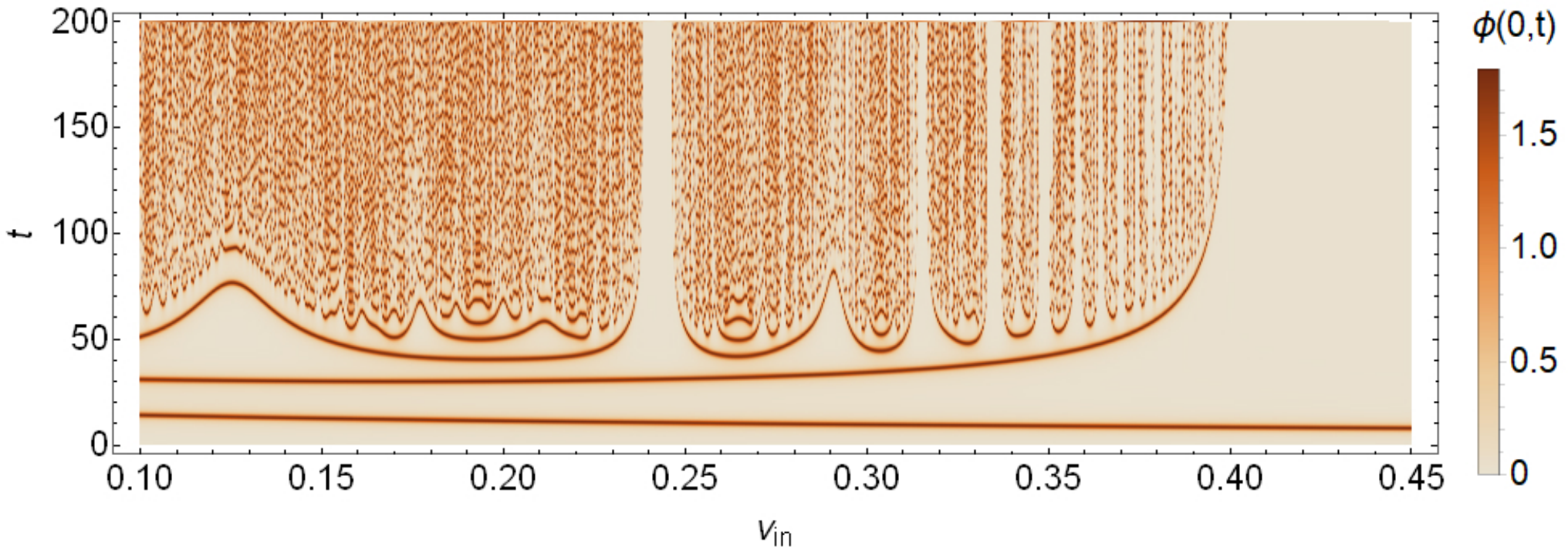}
    \caption{Fractal structure in \added{the} toy model ($\omega_1=2,\,\omega_2=3.46,\,m_{23}=0.11$).}
    \label{fig:4}
\end{figure}

In FIG. \ref{fig:4}, we manifest a set of coefficients that reproduce the fractal structure close to those from the PDE in the $\phi^6$ model.
\section{Conclusion}
%%\label{}
In this work,  we construct\added{ed} a toy model with one translation mode and two oscillation modes, which could reproduce the fractal structure and the MBW phenomena observed in the PDE\added{ simulation} of the $\phi^6$ model. From numerical results, we obtain\added{ed} that our toy model resonance always responds to the lower eigenfrequency\deleted{, which was also observed in \cite{dorey2011kink}}. Furthermore, we \replaced{found}{find} that $m_{23}$, the coupling coefficient between two oscillators, modulate\added{s} the MBWs significantly. 

\replaced{In our toy model, }{Our toy model}\added{we} \replaced{demonstrated}{demonstrates} a possible mechanism of missing bounce windows in soliton dynamics \added{and showed some similarities compared to the full $\phi^6$ model. However, in the full model there are more complex phenomena like spectral walls our toy model cannot reproduce\cite{AdamSpe6,AdamSWSoliton}.}


\bibliographystyle{elsarticle-num}

%\begin{thebibliography}{00}
%\bibliographystyle{unsrt}

%\end{thebibliography}

%% For example:

%% \bibitem[Aladro et al.(2015)]{Aladro15} Aladro, R., Martín, S., Riquelme, D., et al. 2015, \aas, 579, A101

%%\end{thebibliography}

\end{document}